# Correlation of the scaling exponent γ of the diffusivity – density function in viscous liquids with their elastic properties

**Anthony N. Papathanassiou and Ilias Sakellis**

University of Athens, Physics Department, Solid State Physics Section,
Panepistimiopolis, 15784 Zografos, Athens, Greece

**Abstract**

Fundamental thermodynamical concepts and a solid-state point defect elastic model are used to formulate a diffusivity-density scaling function for viscous liquids. It is proved in a straightforward manner that the scaling exponent γ describing the density scaling of the diffusivity, is related with the pressure derivative of the isothermal bulk modulus.

**E-mail addresses:** antpapa@phys.uoa.gr (A.N.P.) and e_sakel@phys.uoa.gr (I.S.)

## I. Introduction

Viscous liquids constitute an exceptional state of matter characterized by extraordinary viscosity values compared with those of ordinary liquids. In the extreme viscosity limit (i.e., close to the calorimetric glass-transition) molecular transport is retarded and most molecular motion is vibrational [1] and the viscous liquid resembles a disordered solid [2]. A smart picture is that of a 'solid that flows' rather than ordinary less viscous liquids [1, 3]. A series of review articles on the properties of glass forming liquids were published recently [1, 2, 3, 4]. Ultra-vviscous matter exhibits many interesting features [5, 6] and universalities which are not well understood yet [2], such as the strongly non-Arrhenius temperature dependence [7] of the structural relaxation time and the strong temperature dependence of the activation energy of the so-called fragile glass formers [1]. A dynamic quantity $\chi$, such as structural relaxation time $\tau$, viscosity $\eta$ or diffusion coefficient D in viscous liquids seems to scale with some fundamental quantities like density $\rho$ and temperature. A popular scaling expression found in the literature [4, 8, 9, 10, 11] is:

$$\chi = F(\rho^\gamma / T) \qquad (1)$$

where $\rho$ denotes the density, $\gamma$ is a scaling exponent, T is the temperature and F is a (scaling) function, which is a priori unknown. Most of the experimental evidence for thermodynamic scaling is for the structural relaxation time and viscosity. Deviations from inverse proportionality between D and $\tau$ occur on approaching the glass transition, whereas enhanced translation relative to reorientation occurs, so scaling of $\tau$ doesn't guarantee scaling of D. However, these deviations from Stokes-Einstein may be small enough to not be apparent in a plot of super positioned data. Or maybe they are subsumed in a small change in $\gamma$ [12]. At this stage, it is abrupt to assume that the diffusivity scaling exponent share a common value with the exponents derived from diffusivity and viscosity experiments. The correlation of the scaling exponent $\gamma$, which is a material constant, with the physical properties of the viscous state is a matter of ongoing exploration. Computer simulations of Lennard-Jones liquids, with the exponent of the repulsive term taking the values 8, 12, 24 and 36, revealed that density scaling is valid and the exponent $\gamma$ is roughly one third of the exponent of the effective inverse power repulsive term [13]. Molecular dynamics also showed that strong virial/potential-energy correlations also reflect the effective inverse power law and scaling occurs in strongly correlating viscous liquids [13]. Recent progress on the role of underlying solid-state point defect elastic models to the density scaling of the diffusivity appeared recently [14]. On the other hand, following the Avramov entropy model [15] for the structural relaxation time, γ was identified to the thermodynamic Grüneisen parameter $\gamma_G$ [4, 11, 16, 17].

The correlation of the scaling exponent γ with some material's constant, most likely with the Grüneisen constant, is the subject of ongoing investigation. Is this the thermodynamic Grüneisen parameter or is it the Grüneisen constant related with a specific process (diffusion, relaxation or viscosity) or some average of different interfering modes? Although the selection of the thermodynamic Grüneisen parameter is justified for ordinary liquids, it is more likely improper to describe the ultra-viscous state of matter, i.e., 'a solid that flows'. This is the reason for turning our attention in correlating γ with some elastic quantity of the viscous liquid that is firmly defined and well determined experimentally. In he present work we formulate a density scaling diffusivity function $D = F(\rho^{((\partial B/\partial P)_T +1)})$, where $(\partial B/\partial P)_T$ is the pressure derivative of the isothermal bulk modulus B. The scaling function implies that the scaling exponent γ is related with the pressure derivative of the isothermal bulk modulus.

**II. Theory, results and discussion**

The isothermal bulk modulus B is defined as:

$$B \equiv -(\partial P/\partial \ln V)_T \qquad (2)$$

Recalling that the density is $\rho \equiv m/V$, we get:

$$B = (\partial P/\partial \ln \rho)_T \qquad (3)$$

To a first approximation, the bulk modulus is assumed to increase linearly with pressure:

$$B(P) = B_0 + (\partial B/\partial P)_T \, P \tag{4}$$

where $B_0$ denotes the zero pressure value of the isothermal bulk modulus and $(\partial B/\partial P)_T$ is assumed to be roughly constant. Volumetric data of various viscous liquids confirm that Eq. (4) is practically a fairly good approximation [18].

Eqs. (3) And (4) merge to:

$$\left(\frac{\partial P}{\partial \ln \rho}\right)_T = B_0 + (\partial B/\partial P)_T \, P \tag{5}$$

By integrating over pressure and density, we get the following equation of state (EOS):

$$\rho^{(\partial B/\partial P)_T} = 1 + \frac{(\partial B/\partial P)_T}{B_0} P \tag{6}$$

where $\rho$ denotes furthermore the reduced density. We note that *the latter EOS is merely based on the condition that B(P) is linear*.

Dielectric measurements in viscous liquids indicate that the logarithm of the relaxation time as a function of pressure (at constant temperature) have a clear non-linear behavior [4, 19, 20, 21]. Information on the dependence of diffusivity upon pressure is merely provided by molecular dynamics simulations [22] in Lennard-Jones mixtures; $\ln D(P)$ isotherms exhibit a downward curvature with respect to pressure. The increase of the (absolute) value of the slope of the latter curve with pressure was speculatively interpreted, as a change in the transport mechanism in viscous liquids, occurring at pressure where hopping of particles become noticeable [22]. Alternatively, it was attributed [22], according to the free-volume theory, to random close packing occurring at elevated pressure.

The activation volume controls the pressure evolution of the diffusivity $\upsilon^{act} \equiv \left(\partial g^{act}/\partial P\right)_T$, where $g^{act}$ denotes the Gibbs free energy for diffusion. Linear lnD(P) plots indicates $\upsilon^{act}$ is constant, while curved ones originate from the pressure dependence of $\upsilon^{act}(P)$ [23]. There is no physical reason to regard $\upsilon^{act}$ as constant; therefore, the compressibility of the activation volume is generally defined as $\kappa^{act} \equiv -\left(\partial \ln \upsilon^{act}/\partial P\right)_T$ [23], and can be positive, negative or zero. The data reported in Ref. [22] indicate that $\kappa^{act} < 0$ for viscous liquids. A diffusing entity can either move in a liquid-like environment by making use of the free volume, or pushing outwards its solid-like environment, or both. No matter what is really the microscopic mechanism, we focused on the thermodynamic quantity of the activation volume, i.e., the volume (density) fluctuation correlated with an activation process. We start from a general diffusivity equation:

$$D(P,T) = f\alpha^2 \nu \exp\left(-g^{act}/kT\right) \qquad (7)$$

where D is the diffusion coefficient, f is a geometrical factor, α is the inter-atomic spacing, ν is the vibrational frequency of the diffusing specie (and related with the phonon frequency involved in the diffusion process) and k is the Boltzmann's constant. Differentiating Eq. (7) with respect to pressure and considering that pressure does not modify the geometrical factor f, we get:

$$\left(\frac{\partial \ln D}{\partial P}\right)_T = -\frac{\upsilon^{act}(P)}{kT} + \frac{1}{B(P)}\left(\gamma_G - \frac{2}{3}\right) \qquad (8)$$

where $\gamma_G$ is the Grüneisen parameter [24]. We further assume that B(P) is linear (Eq. (4)) and that the pressure dependence of the absolute value of the bulk modulus of the activation volume $B^{act} \equiv 1/\kappa^{act}$ is described by the function B(P) governing the bulk volume modification upon pressure. The latter seems quite reasonable; i.e., pressure affects the (absolute value of) activation volume in the same manner pressure reduces the volume of the material. A linear $B^{act}(P)$ implies that:

$$\upsilon^{act}(P) = \upsilon_0^{act}\left[1 + \frac{(\partial B/\partial P)_T}{B_0}P\right]^{1/(\partial B/\partial P)_T} \qquad (9)$$

Therefore, integration of Eq. (8) yields:

$$\ln D(P) = -\frac{\upsilon_0^{act} B_0}{kT(1+(\partial B/\partial P)_T)}\left[\left(1 + \frac{(\partial B/\partial P)_T}{B_0}P\right)^{(1+(\partial B/\partial P)_T^{-1})} - 1\right]$$
$$+ \left(\gamma_G - \frac{2}{3}\right)\frac{1}{(\partial B/\partial P)_T}\ln\left(1 + \frac{(\partial B/\partial P)_T}{B_0}P\right) \qquad (10)$$

where the quantity D(P) is dimensionless and denotes the diffusivity reduced to its zero pressure value $D_0$. It is worth noticing that curved plots are predicted alternatively by assuming that the activation volume is not single valued but obeys the normal distribution [25, 26] By using the EOS described by Eq. (6), Eq. (10) can be expressed in terms of the (reduced) density:

$$\ln D(\rho^{(\partial B/\partial P)_T + 1}) = -\frac{\upsilon_0^{act} B_0}{kT(1+(\partial B/\partial P)_T)}\left[\rho^{((\partial B/\partial P)_T + 1)} - 1\right]$$
$$+ \left(\gamma_G - \frac{2}{3}\right)\frac{1}{((\partial B/\partial P)_T + 1)}\ln\left(\rho^{((\partial B/\partial P)_T + 1)}\right) \qquad (11)$$

Zener [27, 28] asserted that diffusion is controlled by the shear modulus and the Gibbs free energy for migration was set proportional to G. In the 80's, Varotsos and Alexopoulos suggested that the bulk modulus is the elastic quantity tthat controls migration and established proportionality between Gibbs free energy and bulk modulus (cBΩ model) [29 30, 31, 32]. Research on key role of elastic models to understand the peculiar properties of viscous liquids was motivated by Dyre [2, 33]] and still attracts new contributions [34, 35, 36]. Dyre, trying to explain the strong temperature dependence of the activation enthalpy values in viscous liquids, stated that a flow event occurs by re-arrangement of the neighbors of a migrating molecule shoving aside neighboring molecules (shoving model) and, thus, the activation enthalpy is proportional to the shear modulus. However, the question whether shear or bulk elastic moduli control a migration process in an elastic medium is well known to solid-state physicists and was debated during the past decades and, now, reached to an answer: Experimental results for many different types of materials at various experimental conditions (pressure and temperature) support the validity of the cBΩ model. Thus, it seems that the bulk modulus manifests a migration process rather than shear modulus [37, 38]. According to the so-called cBΩ model [29, 30, 31, 32]:

$$g^{act} = cB\Omega \qquad (12)$$

where c is a constant and Ω is a volume related with the mean atomic volume. Note that the validity of Eq. (12) has been checked at ambient pressure in a wide range of solids extending from silver halides [39] to rare gas solids [40], in ionic crystals under gradually increasing uniaxial stress [41] in which electric signals are emitted before fracture (in a similar fashion as the electric signals detected before earthquakes [42, 43, 44, 45], as well as in disordered polycrystalline materials [46]. Differentiating Eq. (12) with respect to pressure we get:

$$\upsilon^{act} = B^{-1}\left[(\partial B/\partial P)_T - 1\right]g^{act} \qquad (13)$$

In the viscous state, the activation enthalpy is usually tenths of kT [2, 47], sometimes even bigger (a range from 60 to 130 kT) was reported [48]. We can write $h^{act} \approx \Lambda kT$. where Λ is a number of the order of 10, which is material dependent [10]. The activation entropy $s^{act}$ is only about a few k, thus, $g^{act} = h^{act} - Ts^{act}$ is of the same order of magnitude as $h^{act}$. Subsequently, at zero pressure, Eq. (13) is rewritten as:

$$\upsilon_0^{act} \approx \frac{\Lambda}{B_0}\left[(\partial B/\partial P)_T - 1\right]kT \qquad (14)$$

Eq. (6), which interconnects density ρ with pressure, and Eq. (14), which links the activation volume with the elastic properties of the material and the large value temperature-dependent activation enthalpy in the viscous matter, transform the generalized diffusivity-pressure relation (Eq. (10)) to a density scaling function:

$$\ln D(\rho^{(\partial B/\partial P)_T+1}) = -\frac{\Lambda((\partial B/\partial P)_T-1)}{((\partial B/\partial P)_T+1)}\left[\rho^{((\partial B/\partial P)_T+1)}-1\right]$$
$$+\left(\gamma_G-\frac{2}{3}\right)\frac{1}{((\partial B/\partial P)_T+1)}\ln\left(\rho^{((\partial B/\partial P)_T+1)}\right) \qquad (15)$$

The latter scaling function indicates that the (scaling) exponent γ governing $\ln D = F(\rho^\gamma)$ is:

$$\gamma = (\partial B/\partial P)_T + 1. \qquad (16)$$

As can be seen in Fig. 1, the morphology of the scaling function is in (qualitatively) agreement with the shape of diffusivity-density scaling plots obtained from simulated Lennard-Jones systems [49, 50, 51]. Moreover, it reproduces recent results indicating that (reduced) diffusivity vs (reduced) density (and temperature) isobars collapse on a master curve [49].

The scaling equation (Eq.(15)) predicts large scaling exponents in relation with those reported from relaxation and viscosity experiments, which vary from 0.1 to 9.0 [4]. Values of $(\partial B/\partial P)_T$ obtained from the analysis of volumetric data of various viscous liquids are given in Ref. [18], which are roughly twice the scaling exponents extracted from relaxation and viscosity data for various representative material classes including Van der Waals liquids, polymers, weakly bonded and ionic materials. Let us consider for example, the case of phenylphyhalein-dimethylether (PDE), for which we have $(\partial B/\partial P)_T = 9.76$ at 362.6 K [18]. By inserting this value into Eq. (16), we find a diffusivity scaling exponent γ=10.76, which is significantly larger than the relaxation scaling exponent 4.5 [50]. In the absence of available experimental diffusivity – density measurements, we can compare with simulation results: molecular dynamics studies, which provide (diffusivity) scaling γ values spanning over a broad range, from 3.5 to 14.5 are reported [49, 51, 52]. It is commonly assumed that the scaling exponents of different dynamic quantities (relaxation, viscosity and diffusivity) of liquids in the viscous state share a common value, which is a material's constant [53]. Coslovich and Roland [54] reported recently simulation results in Kob-Andersen Lennard-Jones mixtures, whereas the relaxation scaling exponent was found to be compatible with the diffusivity exponent found earlier by the same authors for m=12 systems [49]. However, diffusion and relaxation are phenomena of different scale: diffusion is a long-range process, while, relaxation is a short-range one. This idea is inspired from the dielectric response of ionic crystals doped with aliovalent impurities, where the migration enthalpy for vacancy differs from that when the vacancies form (rotating) electric dipoles with aliovalent impurities [55]. The (effective) environment of relaxing entities is modified in comparison with that of a diffusing one. Relaxation and diffusion take place in approximately similar environments below the mode coupling temperature but far above the glass transition, yielding comparable relaxation and diffusivity gammas. However, the difference between the effective potentials, which correspond to each one of the aforementioned mechanisms, by approaching the glass transition, may become more intense and, subsequently, relaxation gammas may diverge each other. Alternatively, the decoupling of translational diffusivity from rotational diffusivity on approaching the glass transition in viscous liquids, is understood as a decoupling of the dynamics occurring on different scales, which arises due to the growing dynamic length scale [56]. Diffusivity experiments would be of great value to provide a full picture of transport in viscous matter, the region close to the glass transition being of great interest for exploration.

## III. Conclusions

The construction of the diffusivity-density scaling function was based on the following:

(i) The equation of state derived from fundamental thermodynamic concepts including the assumption (supported from the experimental data) that the isothermal bulk modulus increases linearly upon increasing pressure.
(ii) Curved diffusivity-pressure isotherms are due to the pressure dependent activation volume controlling the diffusion process.
(iii) The Gibbs free energy for diffusion is proportional to the bulk modulus, which constitutes the so-called cBΩ solid-state point defect model. Elastic models seem to underlie scaling of the dynamic quantities [14, 33, 57, 58].
(iv) The activation energy in fragile liquids is proportional to kT, with a proportionality factor of the order of 10 (or more [48]).

The scaling diffusivity function reproduces qualitatively the plots of diffusivity vs density and temperature. It predicts that diffusivity-density isobars collapse on a common curve, in agreement with recently published simulation results. It is shown in a straightforward manner that the exponent governing scaling of diffusivity in viscous liquids is related with the pressure derivative of the isothermal bulk modulus; i.e., $\gamma = (\partial B/\partial P)_T + 1$.

**Acknowledgements**

Mike Roland (US Naval Laboratory), Daniele Coslovich (Wien Technical University) and Jeppe Dyre (Roskilde University) are greatly acknowledged for their very helpful and informative comments.

24. Briefly, $\gamma_G$ results from the differentiation of the logarithm of the frequency $\nu$ with respect to pressure:

$$\left(\frac{\partial \ln \nu}{\partial P}\right)_T = \left(\frac{\partial \ln \nu}{\partial \ln V}\right)_T \left(\frac{\partial \ln V}{\partial P}\right)_T$$

By definition:

$$\gamma_G \equiv \left(-\frac{\partial \ln \nu}{\partial \ln V}\right)_T$$

and

$$\frac{1}{B(P)} \equiv \left(-\frac{\partial \ln V}{\partial P}\right)_T$$

we get:

$$\left(\frac{\partial \ln \nu}{\partial P}\right)_T = \gamma_G / B(P)$$

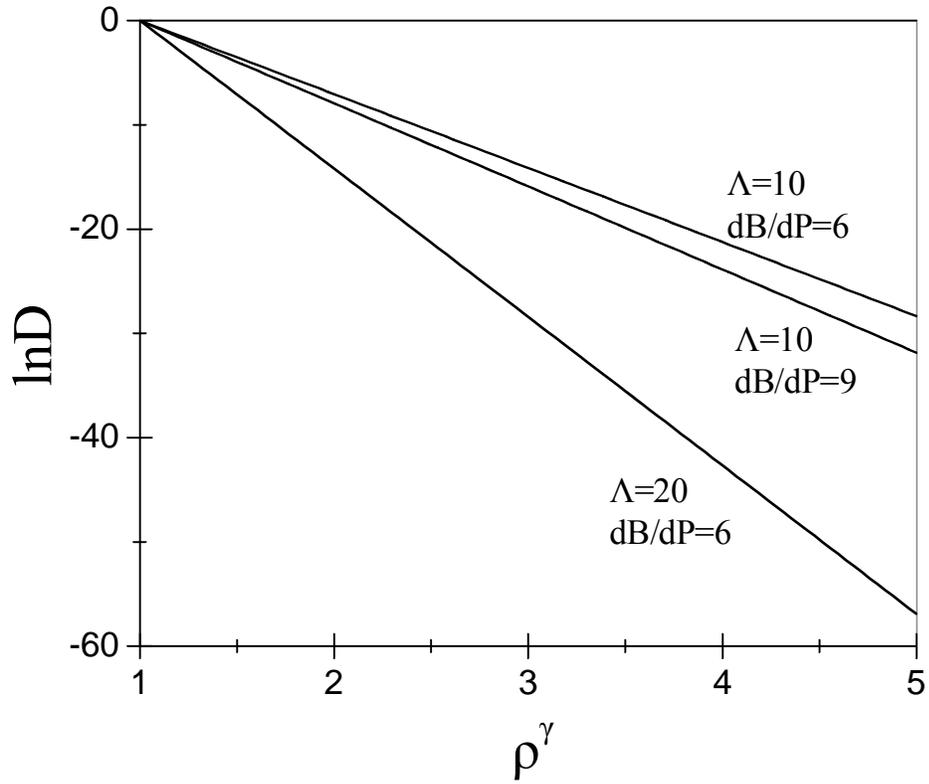

**Figure 1.** Diffusivity plots of the scaling Eq. (15), where $\gamma = (\partial B/\partial P)_T + 1$,, where obtained by regarding different combinations of the parameters $\Lambda$ and $(\partial B/\partial P)_T$. The term $\gamma_G - 2/3$ was taken equal to unity.